# Surveying nearby brown dwarfs with HGCA: direct imaging discovery of a faint, high-mass brown dwarf orbiting HD 176535 A

Yiting Li,[1]★† Timothy D. Brandt,[1]★ G. Mirek Brandt,[1] Qier An,[1]★ Kyle Franson,[2] Trent J. Dupuy,[3] Minghan Chen,[1] Rachel Bowens-Rubin,[4] Briley L. Lewis,[5] Brendan P. Bowler,[2] Aidan Gibbs,[5] Rocio Kiman,[1] Jacqueline Faherty,[6] Thayne Currie,[7] Rebecca Jensen-Clem,[4] Hengyue Zhang,[1,8] Ezequiel Contreras-Martinez,[1] Michael P. Fitzgerald,[5] Benjamin A. Mazin[1] and Maxwell Millar-Blanchaer[1]

[1]*Department of Physics, University of California, Santa Barbara, Santa Barbara, CA 93106, USA*
[2]*Department of Astronomy, The University of Texas at Austin, Austin, TX 78712, USA*
[3]*Institute for Astronomy, University of Edinburgh, Royal Observatory, Blackford Hill, Edinburgh EH9 3HJ, UK*
[4]*Astronomy & Astrophysics Department, University of California, Santa Cruz, CA 95064, USA*
[5]*Department of Physics and Astronomy, University of California, Los Angeles, 475 Portola Plaza, Los Angeles, CA 90025, USA*
[6]*American Museum of Natural History, New York, NY 10023, USA*
[7]*Department of Physics and Astronomy, The University of Texas at San Antonio, San Antonio, TX 78249, USA*
[8]*Sub-department of Astrophysics, Department of Physics, University of Oxford, Keble Road, Oxford OX1 3RH, UK*

## ABSTRACT
Brown dwarfs with well-measured masses, ages, and luminosities provide direct benchmark tests of substellar formation and evolutionary models. We report the first results from a direct imaging survey aiming to find and characterize substellar companions to nearby accelerating stars with the assistance of the Hipparcos–Gaia Catalog of Accelerations (HGCA). In this paper, we present a joint high-contrast imaging and astrometric discovery of a substellar companion to HD 176535 A, a K3.5V main-sequence star aged approximately $3.59^{+0.87}_{-1.15}$ Gyr at a distance of $36.99 \pm 0.03$ pc. In advance of our high-contrast imaging observations, we combined precision High Accuracy Radial velocity Planet Searcher (HARPS) Radial Velocities (RVs) and HGCA astrometry to predict the potential companion's location and mass. We thereafter acquired two nights of KeckAO/NIRC2 direct imaging observations in the $L'$ band, which revealed a companion with a contrast of $\Delta L'_p = 9.20 \pm 0.06$ mag at a projected separation of $\approx 0.35$ arcsec ($\approx 13$ au) from the host star. We revise our orbital fit by incorporating our dual-epoch relative astrometry using the open-source Markov chain Monte Carlo orbit fitting code ORVARA. We obtain a dynamical mass of $65.9^{+2.0}_{-1.7} M_{\rm Jup}$ that places HD 176535 B firmly in the brown dwarf regime. HD 176535 B is a new benchmark dwarf useful for constraining the evolutionary and atmospheric models of high-mass brown dwarfs. We found a luminosity of $\log(L_{\rm bol}/L_{\odot}) = -5.26 \pm 0.07$ and a model-dependent effective temperature of $980 \pm 35$ K for HD 176535 B. We infer HD 176535 B to be a T dwarf from its mass, age, and luminosity. Our dynamical mass suggests that some substellar evolutionary models may be underestimating luminosity for high-mass T dwarfs. Given its angular separation and luminosity, HD 176535 B would make a promising candidate for Aperture Masking Interferometry with *JWST* and GRAVITY/Keck Planet Imager and Characterizer, and further spectroscopic characterization with instruments like the CHARIS/SCExAO/Subaru integral field spectrograph.

**Key words:** methods: data analysis – techniques: high angular resolution – techniques: image processing – astrometry.

## 1 INTRODUCTION

Brown dwarfs (BDs) are substellar objects above the deuterium burning mass limit ($\geq \sim 13\, M_{\rm Jup}$; Luhman 2008) but with insufficient mass ($\leq \sim 75\, M_{\rm Jup}$) to achieve the core densities and pressures to sustain stable nuclear fusion of hydrogen to helium (Hayashi & Nakano 1963; Kumar 1963). BDs may form isolated, in multiple systems or in protostellar discs and envelopes around single stars (Chabrier et al. 2000; Stamatellos, Hubber & Whitworth 2007; Basu & Vorobyov 2012; Thies et al. 2015). After their formation, they gradually lose energy radiatively and become denser and cooler with time, until electron degeneracy pressure halts gravitational collapse (Whitworth 2018). As they age and cool, they evolve across the M, L, T, and Y spectral types categorizable with infrared spectroscopy. For example, an L-dwarf ($T_{\rm eff} = \sim 1250$–$2000$ K) spectrum is dominated by metal hydride absorption bands (FeH, CrH, MgH, CaH) and prominent alkali metal lines (Na I, K I, Cs I, Rb I), while the cooler

★ E-mails: yitingli@ucsb.edu (LY); tbrandt@physics.ucsb.edu (BT); aan@ucsb.edu (QA)
† NSF Graduate Research Fellow.

T dwarfs with surface temperatures of ∼500–1250 K have methane-rich atmospheres characterized by absorption lines of $H_2O$, CO, TiO, and $CH_4$ (Hubbard 2000; Geballe et al. 2001; Lodders & Fegley 2006; Delorme et al. 2008; Phillips et al. 2019). The chemically diverse atmospheres of BDs disclose crucial information on their appearance, physical processes, evolution, cooling, effective temperature, and luminosity (Madhusudhan, Apai & Gandhi 2016).

Over the last two decades, the evolution of BDs became mainly understood through theoretical modelling (Marley et al. 1996; Baraffe et al. 2002; Saumon & Marley 2008; Marley & Robinson 2015). In particular, cloud formation in BD atmospheres is thought to be an important factor in understanding major variations in their thermal spectra and observable properties, especially those on the L–T transition (Cooper et al. 2003; Morley et al. 2014). BDs cool on a time-scale of several million years post-formation, and their cooling follows a mass–luminosity–age relation. Uncertainties in the measurements of mass, age, or luminosity, especially in mass and age, present a major challenge for using them to test and calibrate evolutionary and BD cooling models (Burrows, Hubbard & Lunine 1989; Liu et al. 2009). Therefore, BDs whose properties are measured independent of evolutionary models serve as important benchmark objects anchoring our understanding of BD properties and their evolution.

Young and/or massive BDs may be directly imaged with the high-contrast imaging technique, providing measurements of their spectra and luminosities. Furthermore, those with RV curves may also be amenable to dynamical mass measurements. To date, close to 30 BDs with a range of ages and orbital periods have been directly imaged and characterized. These BDs have been routinely used as benchmarks for substellar evolutionary models (Crepp et al. 2013, 2016, 2018; Dupuy & Liu 2017; Brandt et al. 2020, 2021d; Currie et al. 2020, 2022; Bowler et al. 2021b; Chen et al. 2022; Franson et al. 2022; Kuzuhara et al. 2022; Rickman et al. 2022; Swimmer et al. 2022). The T-type BDs with spectral types T0–T9 were first observed and classified in 1995 (Oppenheimer et al. 1995), which are cool BDs exhibiting methane absorption features in the near-infrared wavelengths of 1–2.5 μm, distinguishing them from the warmer L dwarfs. Most measured T dwarfs have relatively low masses (∼30–60$M_{\rm Jup}$) given their low temperatures, allowing them to progressively cool through the T dwarf sequence over 1–5 Gyr.

With more and more benchmark T dwarfs being observed and characterized, there emerges a small but alarming list of overmassive and underluminous T dwarfs that challenge the validity of most substellar cooling models, which predict that such massive objects cannot sufficiently cool to their observed low temperatures within their (approximately) known ages. For example, the first T dwarf Gl 229 B's very high mass ($71.4 \pm 0.6 M_{\rm Jup}$) and low luminosity do not match with the model predictions even at a very old age (Brandt et al. 2020, 2021d). The late-T dwarf HD 4113 C is another example presenting a severe challenge to evolutionary models because its measured dynamical mass of $66 M_{\rm Jup}$ is significantly higher than the model prediction of ∼$40 M_{\rm Jup}$ given its low luminosity of $10^{-6}$ L$_\odot$ (Cheetham et al. 2018). The late-T dwarf HD 47 127 B is a massive and old (7–10 Gyr) BD companion to a white dwarf, and another peculiar case where the models prefer a high mass of $100 M_{\rm Jup}$ that is at odds with the measured mass of 67–78 $M_{\rm Jup}$ (Bowler et al. 2021b).

However, despite the couple of outliers, most benchmark T dwarfs appear to be in agreement with evolutionary models (e.g. Dupuy & Liu 2017; Brandt, Dupuy & Bowler 2019; Brandt et al. 2021d): both T dwarfs of the binary BD system ε Indi B were previously believed overmassive T dwarf candidates (Dieterich et al. 2018), but the recent reanalysis by Chen et al. (2022) supports the consistency of evolutionary models. One possible resolution to the overmassive T dwarf problem is unresolved multiplicity. In this scenario, nearly all of the light would be due to the observed T dwarf, while a modest fraction of the mass would be hidden in a somewhat lower mass, but much fainter, companion. Unresolved, lower mass companions would contribute almost nothing to the overall integrated spectra of the BDs at older ages. If binarity can be ruled out as the cause of the disagreement, these overmassive BDs may point to a strong tension with evolutionary models. A larger sample of well-characterized BDs will be required in order to investigate the T dwarf population.

Astrometry has recently emerged as a way to identify stars being tugged across the sky by massive, unseen companions, a new method to identify hosts of T dwarfs. Satellites such as ESA's *Hipparcos* and *Gaia* have provided absolute astrometry for billions of stars across the sky; the Hipparcos–Gaia Catalog of Accelerations (HGCA; Brandt 2021) has cross-calibrated *Hipparcos* and *Gaia* astrometry. Here, we report the initial findings from a pilot survey utilizing the Near-Infrared Camera 2 (NIRC2) vortex coronagraph, operating in the L'-band with the Adaptive Optics (AO) system of the Keck Observatory. Our survey focuses on the detection of substellar companions around stars exhibiting astrometric accelerations in the HGCA. With astrometry, one can identify targets worthy of following up by deriving their orbits from RV and absolute astrometry joint fits. Also, using this new technique to pre-identify targets, several groups have carried out similar follow-up imaging surveys (Fontanive et al. 2019; Currie et al. 2021; Bonavita et al. 2022; Franson et al. 2023) that greatly improved detection rates unparalleled by traditional blind imaging surveys. The T dwarf companion HD 176 535 B presented in this paper is among the first companions discovered whose astrometric location was known before imaging. In this paper, we present an ORVARA fit incorporating our new Keck/NIRC2 astrometry data to constrain the mass and three-dimensional (3D) orbit of the system.

We construct the paper as follows. We review the host star's stellar characteristics in Section 2. In Section 3, we describe our input data for our orbital fit, including archival High Accuracy Radial velocity Planet Searcher (HARPS) RV data, the proper motions from the HGCA, and our new relative astrometry from Keck/NIRC2 AO imaging observations. Section 4 presents our joint orbital fit and results. We benchmark HD 176 535 B against evolutionary models and discuss its implications in Section 5. Finally, we conclude our findings with Section 6.

## 2 STELLAR PROPERTIES

HD 176 535 is a main-sequence K3.5 V (Gondoin 2020) star located at a distance of $36.99 \pm 0.03$ pc based on *Gaia* EDR3 (Gaia Collaboration 2020). HD 176535's stellar properties are similar to those of the Sun in terms of age, metallicity, kinematics, and atmospheric chemical abundances; it has a history of being identified as a solar sibling candidate that may share a common birth cluster with the Sun (Batista et al. 2014; Adibekyan et al. 2018). HD 176 535 has an effective temperature of $4727 \pm 104$ K (Sousa et al. 2011) and a surface gravity of $4.63 \pm 0.05$ dex (Gaia Collaboration 2022). HD 176 535 is slightly metal poor with an iron abundance of [Fe/H] ≈ −0.15 dex (Sousa et al. 2011; Adibekyan et al. 2012; Mortier et al. 2013; Gáspár, Rieke & Ballering 2016; Suárez-Andrés et al. 2017; Gondoin 2020). The chromospheric index of $\log_{10} R'_{\rm HK} \approx -4.732$ dex (Gondoin 2020) indicates that HD 176 535 is an old and relatively inactive star. We summarize the properties of the HD 176 535 AB system in Table 1.

HD 176535's age was determined on several accounts. While analysing a sample of HARPS FGK stars using *Gaia* DR2 parallax and the PAdova and TRieste Stellar Evolution Code (PARSEC) (Bressan et al. 2012), Delgado Mena et al. (2019) derive an isochronal age of 5.21 ± 4.68 Gyr for HD 176 535. Besides, Gaia Collaboration (2022) gives an age of 3.34 Gyr using the Fitting Location of Age and Mass with Evolution (FLAME) model with 1σ lower and upper confidence levels of 0.61 and 6.97 Gyr, respectively. The FLAME model uses the PARSEC isochrones to generate model predictions and estimate the age and mass of stars based on their observed properties. Gomes da Silva et al. (2021) found a chromospheric index of $\log_{10} R'_{HK} = -4.5695 \pm 0.0069$ dex by measuring the Ca II HK emission lines in the star's HARPS spectra, and thus found HD 176 535 A to be a relatively old star with a weakly constrained age of 5.57 ± 4.84 Gyr. Delgado Mena et al. (2019) analysed the same HARPS spectra as did Gomes da Silva et al. (2021), but found a different S-index of 0.43 rather than Delgado Mena et al. (2019)'s value of 0.63. We perform our own independent analysis to constrain the age of HD 176 535 A using a Bayesian-based age-dating method developed by Brandt et al. (2014). We get an age of $5.02^{+1.11}_{-1.39}$ Gyr if adopting only the S-index of 0.43 from Pace, and a value of $2.57^{+0.67}_{-0.89}$ Gyr if adopting the S-index of 0.63 from Gomes. If otherwise adopting both S-indices as lower and upper limits of $\log_{10} R'_{HK}$, we retrieve a somewhat young age of $3.59^{+0.87}_{-1.15}$ Gyr for HD 176 535 A. Although our age estimates are broadly consistent with literature measures, none of the ages were well established and the age of the system remains largely ambiguous. The rotation period of the star has neither been reported in the literature nor observed in any *TESS* sector. Future light curves from *TESS*, when observed, would enable retrieval of a rotation period and thus increase the confidence in the system's age.

The mass of the star is estimated to be between 0.69 and 0.76 $M_\odot$ via several literature sources including the *Gaia* DR3 catalogue (Pinheiro et al. 2014; Chandler, McDonald & Kane 2016; Anders et al. 2019; Delgado Mena et al. 2019; Stassun et al. 2019; Gomes da Silva et al. 2021; Gaia Collaboration 2022). In the same way as described in Li et al. (2021), we obtain an isochronal mass of 0.72 ± 0.02 $M_\odot$ using the PARSEC model. For this work, we adopt our estimated age of $3.59^{+0.87}_{-1.15}$ Gyr and our estimated mass of 0.72 ± 0.02 $M_\odot$ for the host star HD 176535.

## 3 DATA AND OBSERVATIONS

### 3.1 Archival RV data

RVs of HD 176 535 come from the HARPS spectrograph mounted at the ESO 3.6-m telescope in La Silla. HARPS is a state-of-the-art fibre-fed high-resolution ($R \sim 115\,000$) optical Echelle spectrograph that achieves sub-m/s RV precision (Mayor et al. 2003). We retrieve 18 RVs for HD 176 535 from the HARPS public radial velocity data base, which has been corrected for common RV systematic errors by Trifonov et al. (2020). These RVs were taken between 2004 and 2015 with a median uncertainty of 1.46 m s$^{-1}$. All were taken prior to the 2015 HARPS fibre upgrade. HD 176 535 A is a K star with relatively low level of stellar activity, allowing a high degree of RV precision.

### 3.2 *Hipparcos–Gaia* accelerations

*Hipparcos* and *Gaia* absolute astrometry have enabled precise dynamical mass measurements of both RV-discovered and directly imaged companions (Brandt et al. 2019, 2021a; Currie et al. 2020, 2022; Bowler et al. 2021a; Li et al. 2021; Franson et al. 2022; Kuzuhara et al. 2022). Absolute astrometry provides the tangential component of the host star's acceleration, complementing both stellar radial velocities and relative astrometry that constrains the

**Table 1.** Properties of the HD 176 535 AB system.

| Property | Value | Refs. |
|---|---|---|
| Host star | | |
| $\varpi$ (mas) | 27.033 ± 0.018 | 1 |
| Distance (pc) | 36.99 ± 0.03 | 1 |
| SpT | K3.5V | 2, 3 |
| Mass ($M_\odot$) | 0.72 ± 0.02 | 4, 5 |
| Age (Gyr) | 5.04 ± 1.45 | 12 |
| $T_{\text{eff}}$ (K) | 4727 ± 104 | 6 |
| [Fe/H] (dex) | −0.15 ± 0.07 | 7 |
| $\log(R'_{HK})$ (dex) | −4.85 | 8 |
| $R'_X$ (dex) | <−4.28 | 9 |
| *Gaia* RUWE | 1.019 | 1 |
| Luminosity ($L_\odot$) | 0.208 ± 0.007 | 12 |
| *Gaia* G (mag) | 9.374 ± 0.003 | 1 |
| $B_T$ (mag) | 11.195 ± 0.066 | 10 |
| $V_T$ (mag) | 9.923 ± 0.035 | 10 |
| J (mag) | 7.804 ± 0.027 | 11 |
| H (mag) | 7.313 ± 0.033 | 11 |
| $K_s$ (mag) | 7.175 ± 0.020 | 11 |
| Companion | | |
| Mass ($M_{\text{Jup}}$) | $65.9^{+2.0}_{-1.7}$ | 12 |
| $L'$ apparent (mag) | 16.31 ± 0.07 | 12 |
| $L'$ absolute (mag) | 13.47 ± 0.07 | 12 |
| Semimajor axis (au) | $11.05^{+0.64}_{-0.56}$ | 12 |
| Inclination (°) | $49.8^{+3.4}_{-3.7}$ | 12 |
| Period (yr) | $40.6^{+3.9}_{-3.5}$ | 12 |
| Eccentricity | $0.496^{+0.022}_{-0.020}$ | 12 |

*Note.* References abbreviated as (1) Gaia Collaboration (2020); (2) Gray et al. (2006); (3) Bourgés et al. (2014); (4) Reiners & Zechmeister (2020); (5) Delgado Mena et al. (2019); (6) Sousa et al. (2011); (7) Gáspár et al. (2016); (8) Pace (2013); (9) Voges et al. (1999); (10) Høg et al. (2000); (11) Cutri et al. (2003); and (12) this work.

orbital motion of the companion. The proper motions measured by *Hipparcos* and *Gaia*, along with the scaled positional difference over the 25-yr baseline between these two proper motions, allow precise constraints on a companion's 3D orbit. Changes in these proper motions indicate acceleration of the host star in an inertial reference frame. In order to compare the proper motions from *Hipparcos* and *Gaia*, Brandt (2018, 2021) carried out a cross-calibration between the two catalogues by establishing a common reference frame and correcting for error inflation; this is presented in the HGCA.

HD 176 535 A is catalogued in the HGCA as a high-proper-motion star observed by both *Hipparcos* and *Gaia*. *Gaia* EDR3 itself provides the most precise proper motion measurement. The long-term proper motion – the difference in position between *Hipparcos* and *Gaia* scaled by their time baseline – is almost as precise. These two proper motions differ, demonstrating a significant acceleration of ~90σ between the long-term *Hipparcos*–*Gaia* proper motion and *Gaia* (EDR3). A summary of the absolute astrometry from the HGCA for HD 176 535 A is given in Table 2. HD 176 535 has a renormalized unit weight error, or RUWE, of 1.019 in *Gaia* EDR3. This is close to 1 and significantly less than 1.4, indicating that a five-parameter astrometric model (with position, proper motion, and parallax) provides a good fit to the data (Gaia Collaboration 2020; Stassun & Torres 2021). The low RUWE strongly disfavours a close, massive companion with a period close to the 33-month baseline of *Gaia* EDR3.

We perform a preliminary orbital analysis using the existing HARPS RVs and *Hipparcos*–*Gaia* absolute astrometry. This enables evaluation of the detectability of the companion, as well as

**Table 2.** HGCA absolute astrometry for HD 176 535 A.

| Parameter | *Hipparcos* | *Hipparcos–Gaia* | *Gaia* EDR3 |
| --- | --- | --- | --- |
| $\mu_{\alpha*}$ (mas yr$^{-1}$) | −21.9 ± 1.3 | −22.016 ± 0.048 | −24.927 ± 0.025 |
| $\mu_\delta$ (mas yr$^{-1}$) | −32.1 ± 1.0 | −32.295 ± 0.030 | −29.707 ± 0.022 |
| corr ($\mu_{\alpha*}, \mu_\delta$) | 0.13 | −0.02 | 0.21 |
| $t_\alpha$ (Jyr) | 1991.27 | – | 2016.27 |
| $t_\delta$ (Jyr) | 1991.41 | – | 2016.36 |

*Note.* The $\chi^2$ value for a model of constant proper motion (*Hipparcos–Gaia* and *Gaia* proper motions are equal) is 8215 with 2 degrees of freedom.

predictions for the location of the companion prior to any on-sky observation. We use the Markov chain Monte Carlo (MCMC) orbit code ORVARA (Brandt et al. 2021b) to jointly fit the HARPS RVs and HGCA stellar astrometry. Our RV and absolute astrometry-only fit reveals the existence of an ≈63 $M_{\rm Jup}$ companion at a distance of ≈14 au in orbit around HD 176535A. However, for RV systems, even with the addition of complementary HGCA absolute astrometry data, the orbital inclination of the system may still not be precisely determined due to the uncertain direction of motion of the companions (Li et al. 2021). For such systems, their orbits can be resolved either by future *Gaia data* releases or by high-contrast imaging.

ORVARA is capable of generating 3$\sigma$ likelihood contours for the coordinates of the companion with respect to its host star at an arbitrary epoch. Our preliminary fit shows that HD 176 535 AB is an excellent system for imaging follow-up. The broad inclination posterior distributions give out a range of possibilities for the location of the companion with respect to the host star. More detailed use of ORVARA will be discussed in Section 4. We display the predicted contours that trace out the possible locations of the companion in the left-hand panel of Fig. 1 for our first observation of the system in August 2021.

### 3.3 Keck/NIRC2 AO imaging

Our high-contrast imaging observations of HD 176 535 took place on both UT 2021 August 26 and UT 2022 July 7 in the $L'$ band (3.426–4.126 μm) using the NIRC2 camera in pupil-tracking mode at the W. M. Keck Observatory (WMKO) with the Vector Vortex Coronagraph (Serabyn et al. 2017). The observations were carried out with natural guide star adaptive optics (Wizinowich 2013) and the visible-light Pyramid Wavefront Sensor (WFS; Bond et al. 2020). The observations were set to track the telescope pupil, allowing the sky to rotate during the observation. We use the standard set-up for vertical angle mode to track the rotating pupil images when carrying out angular differential imaging (ADI; Marois et al. 2006). The ADI images of the HD 176535AB system were taken in sequences of 30 science frames using the Quadrant Analysis of Coronagraphic Images for Tip–tilt Sensing (QACITS) algorithm, which centres the star behind the vortex phase mask via small tip–tilt corrections after each exposure (Huby et al. 2015). Each QACITS sequence includes off-axis unsaturated images for flux calibration and optimization and sky background frames for the science and calibration images. For both of our nights, we observed with 60 coadded integrations and 0.5-s integration time per coadd. We obtain a total of 90 ADI images for the August 2021 observation with a total integration time of 45 min and 79 images for the July 2022 night with a total integration time of 39.5 min. In addition, for each of the two nights, dark and twilight flat frames were taken at the end of the night for image reduction.

We employ the Vortex Image Processing (VIP) package (Gomez Gonzalez et al. 2017) for post-processing. We stack the images for HD 176535AB into 3D ADI sequences and reduce the 3D data cube with the following procedure. First, cosmic rays are removed using the lacosmic PYTHON package (van Dokkum 2001) and geometric correction was performed on each ADI cube with the solutions found by Service et al. (2016) for the narrow-field mode of the NIRC2 camera. We perform sky subtraction for the science and the off-axis flux calibration frames using the AstroDrizzle sky-subtraction function from Avila et al. (2015). We then subtract the dark frames and flat-field all the images, and correct for bad pixels. The centring and alignment of the raw ADI images at subpixel accuracy is a crucial pre-processing step that directly affects the quality of data reduction. Thus, we use VIP to perform re-centring, rotation, and realignment of the images by fitting negative two-dimensional (2D) Gaussian profiles to the vortex core in each time slice.

After image re-centring is complete, we use VIP to perform stellar Point Spread Function (PSF) subtraction. The stellar PSF is obtained by subtracting the sky frame from the PSF frame with the star off the vortex. Through fitting 2D Gaussians to the host star's PSF, we measure a full width at half-maximum (FWHM) of 7.80 pixels (0.078 arcsec) for both epochs. VIP provides several ADI-based algorithms to handle model PSF subtraction, including Median Combination of Images, the Locally Optimized Combination of Images (Lafrenière et al. 2007), annular and full-frame Principal Component Analysis (PCA), PCA in a single annulus, annular and full-frame Non-negative matrix factorization low-rank approximation (Gomez Gonzalez et al. 2017; Ren et al. 2018), and Local Low-rank plus Sparse plus Gaussian-noise decomposition (Gomez Gonzalez et al. 2016). We choose PCA-based algorithms (Amara & Quanz 2012; Soummer, Pueyo & Larkin 2012) for the best performance in speed and efficiency. In ADI data cubes, any putative planet would rotate throughout the images with time, while the stellar halo and speckle pattern stays quasi-static. A reference PSF must be constructed for subtraction to create a residual cube. In PCA-based algorithms, this reference PSF is constructed from the projection of each image on a subset of the principal components (PCs). Too few components result in incomplete removal of the stellar PSF, while too many lead to oversubtraction of the companion, with the companion increasingly present in the PCs. The PCs are computed through singular value decomposition of the 2D matrix (time × flattened image array) of the images. The residual cube is then derotated to align the field of view and coadded to create a final image.

We apply annular PCA from VIP to do the PSF subtraction, which uses the PCA low-rank approximation annulus-wise to capture the background and speckle noise for a given concentric annulus (Gomez Gonzalez et al. 2016). In annular PCA, a frame rejection criterion (Absil et al. 2013) can be applied based on a parallactic angle (PA) threshold. By keeping only the frames where the planet has rotated by this threshold, we can prevent the planetary signal from being captured by the low-rank approximation and thus subtracted from the science images. We tune the amount of rotation gap used with a PA threshold from 0.1× FWHM to 1× FWHM, and determined a minimum PA rotation between images of $\delta_{\rm rot} = 0.3 \times$ FWHM at the companion's location that maximizes the signal-to-noise ratio (SNR) in the final frame. We optimize the SNR for each individual companion at the same time by adjusting the number of PCs. We run PSF subtraction for the pre-processed images varying the number of components from 0 to 40, each time measuring the SNR at the location of the companion. For each number of PCs used, the signal and noise levels are calculated with relative photometry using FWHM-radius circular apertures. The relative astrometry of the companion each time is determined by fitting a simple 2D Gaussian to a 15 × 15 pixel subimage. The SNR is computed as the ratio of the companion flux and the standard deviation of the flux within the noise estimation apertures via the method outlined by Mawet et al. (2014). Fig. 2 shows the SNR as a function of the number of PCs used in annular PCA analysis for both epochs. The optimal number of PCs that maximize SNR is 6 for the 2021 epoch with an SNR of 11.7, and 3 components for the 2022 epoch with an SNR of 10.2. The subtracted images are shown in Fig. 1 where the companion is clearly visible east of the host star.

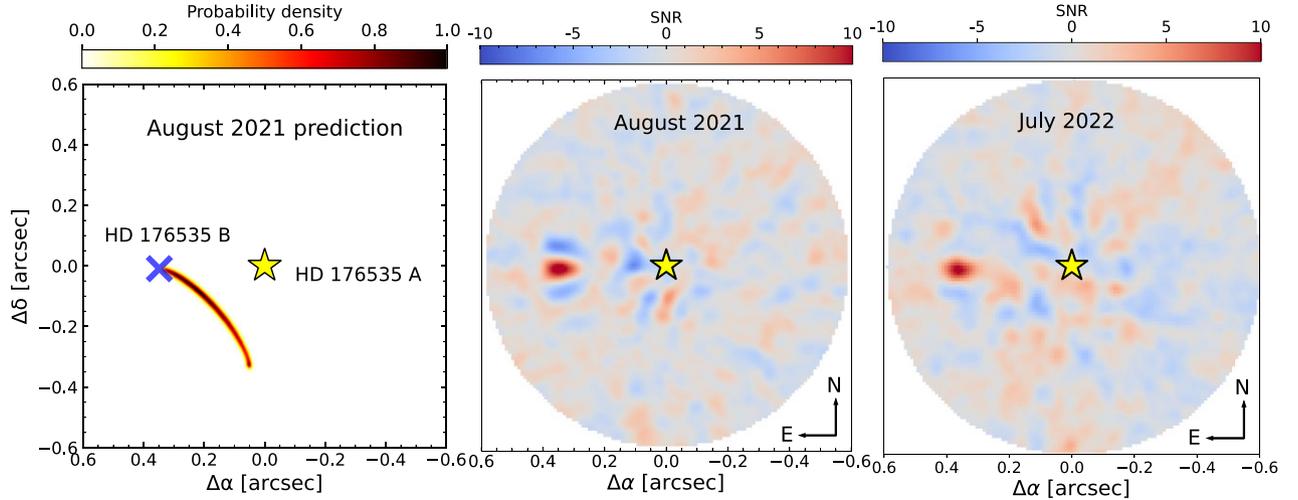

**Figure 1.** (Left) Predicted location of HD 176 535 B is shown by the $3\sigma$ probability contour before its first imaging in August 2021. The blue cross is the true location of the companion where it was imaged on 2021 August 26. (Middle and right) PSF-subtracted images of HD 176 535 B for the August 2021 and July 2022 detections. The yellow star in the centre is the host star's location, and the companion was detected both times to the east of the host star. The images are smoothed with a Gaussian filter with a window function of the form $e^{-(x^2+y^2)}$.

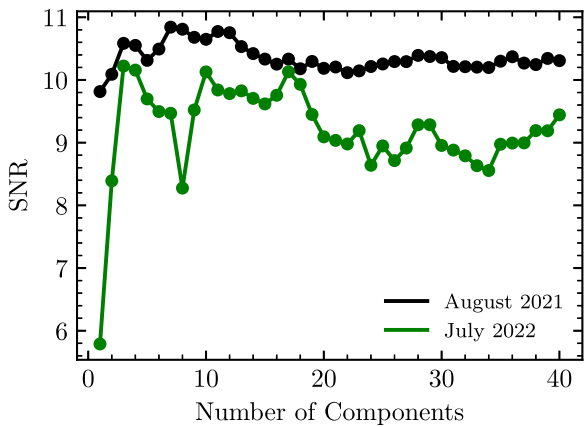

**Figure 2.** The SNR of HD 176 535 B in the PSF-subtracted Keck/NIRC2 images as a function of the number of annular PCA components for both August 2021 and July 2022 images. The optimal number of components that yield the best signal is 6 components for August 2021 with an SNR of 11.7, and 3 components for July 2022 with an SNR of 10.2. We adopt these values to produce our annular PCA PSF subtractions.

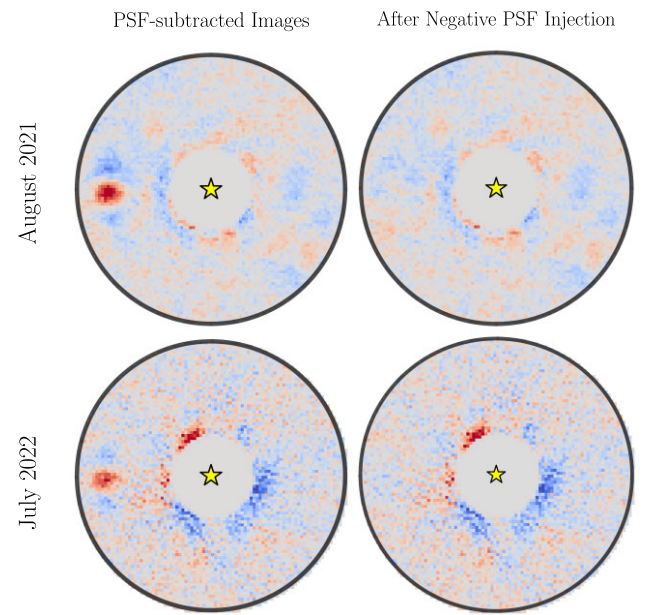

**Figure 3.** The images before (left) and after (right) injecting negative PSF templates with the best-fitting separations, position angles, and flux ratios. The left-hand panels show the PSF-subtracted images of HD 176 535 B from the August 2021 and July 2022 Keck/NIRC2 data sets. The right-hand panels are the results after NEGFC injection. The signals of the companions are completely removed after the negative PSF injection. The increase in the noise level near the centre is due to proximity to the stellar halo, resulting in poor PSF subtraction.

In stellar PSF subtraction approaches, a 2D Gaussian fit leads to biased results for the astrometry and photometry of a companion because part of the planetary signal is self-subtracted during the stellar halo removal process. To avoid systematic biases, we adopt the negative fake companion (NEGFC) technique described in Lagrange et al. (2010), Marois, Macintosh & Véran (2010), and Wertz et al. (2017) for robust extraction of the position and flux of detected point-like sources. This method involves the injection of a negative-amplitude PSF template at the location of the companion. Ideally, the injection of a negative-flux synthetic companion completely removes the signal from the true companion in the final image. The NEGFC method proceeds as follows. First, one obtains an estimate of a 'first guess' for the (biased) position and flux of the companion from the first image of the reduced ADI data cube. Then, this measured normalized off-axis PSF is scaled and shifted to remove the first estimate from the input data cube before applying PCA. Finally, one applies PCA in a single (optimal) annulus and iterates on the position and flux of the injected negative PSF template, until the absolute residuals $\chi^2$ in the aperture are minimized (i.e. when the injected negative companion flux and position best match those of the true companion). To do the iteration from the last step, we first use a Nelder–Mead simplex minimization algorithm to get a starting guess for the flux and position of the true companion. We then use the EMCEE affine-invariant MCMC ensemble sampler (Foreman-Mackey et al. 2013) to obtain the companion's flux (flux$_c$) and angular coordinates ($\rho$, $\theta$). With MCMC, we can infer the highest likelihood parameter values and uncertainties by sampling the posterior distributions of the parameters ($\rho$, $\theta$, flux$_c$). We use 100 walkers for our MCMC chain, and iterate over $10^5$ steps until convergence is achieved. We use the autocorrelation time-based criterion $N/\tau \geq a_c$ with $a_c = 50$ (Christiaens et al. 2021) to test convergence. We discard 30 per cent of the chain as burn-in. Fig. 3 displays the PSF-subtracted images and the results following the injection of negative fake companions for each epoch.

We measure the $L'$-band contrast for the HD 176 535 AB system by separately measuring the photometry for HD 176 535 A and the NEGFC of HD 176 535 B from MCMC for both August 2021 and July 2022 NIRC2 data sets. We scale the flux ratio according to exposure times. Ultimately, for July 2021 data, we measure a contrast of $\Delta L' = 9.19 \pm 0.07$ mag. We estimate the star's $L'$-band magnitude from both its *WISE* and from its near-infrared photometry. HD 176 535 A has a *WISE* W1 magnitude of $7.11 \pm 0.02$ mag (Maldonado et al. 2017). Since the $L'$ band is on HD 176 535 A's Rayleigh–Jeans tail and has a central wavelength nearly identical to that of *WISE* W1, the star's W1 and $L'$ magnitudes are nearly identical. We also transform from 2MASS $H$ ($7.313 \pm 0.033$ mag; Cutri et al. 2003) to $K$ band and subsequently to $L'$ using the relations of Bessell & Brett (1988). This approach gives the an $L'$ magnitude of $7.11 \pm 0.04$, identical to the $L'$ photometry inferred from W1 albeit with a larger uncertainty. We adopt this larger uncertainty for a final $L'$-band magnitude of $L'_* = 7.11 \pm 0.04$ mag. This stellar magnitude and contrast then produce a magnitude of $L'_p = 16.30 \pm 0.07$ for the companion HD 176 535 B. Likewise, for the July 2022 NIRC2 data set, we measure a contrast of $\Delta L' = 9.21 \pm 0.10$ and an $L'$ magnitude of $L'_p = 16.32 \pm 0.10$ for HD 176 535 B. We combine the two epochs to get a final contrast of $\Delta L' = 9.20 \pm 0.06$. We then add the star's magnitude of $L' = 7.11 \pm 0.04$ to obtain an apparent $L'$ magnitude for the companion of $16.31 \pm 0.07$ or an absolute magnitude of $13.47 \pm 0.07$.

The uncertainties in our NIRC2 astrometry are dominated by four error terms: the uncertainties in the measurement of the astrometry of the companion from the MCMC posteriors, the distortion correction to the NIRC2 narrow camera individual frames, the NIRC2 plate scale, and the north alignment of the detector. The raw images are corrected for optical geometric distortion using the post-2015 solution from Service et al. (2016) after the adaptive optics system and the NIRC2 camera were realigned. They found a post-realignment residual distortion solution through the mean rms scatter in the position measurements of dozens of stars, which is about 1.1 mas in both $X$ and $Y$. We adopt a post-alignment global plate scale of $9.971 \pm 0.004 \pm 0.001$ mas per pixel for the Keck/NIRC2 (Marois et al. 2008; Bowler et al. 2012; Serabyn et al. 2017; Xuan et al. 2018) AO system in narrow mode (Service et al. 2016). For the position angle, Service et al. (2016) give an angle of $\theta_{\rm north} = 0°.262 \pm 0°.020$ to align the images with celestial north. Thus in summary, the position angle of the companion must be corrected from the uncorrected MCMC measurements of the position angle $\theta_{\rm meas}$ through the following equation:

$$\theta = \theta_{\rm meas} - {\rm PARANG} - {\rm ROTPOSN} + {\rm INSTANGL} - \theta_{\rm north}, \quad (1)$$

where PARANG is the PA, ROTPOSN is the rotator position of $4°.43$, and INSTANGL is the NIRC2 position angle zero-point of $0°.7$. Table 3 showcases the dual-epoch relative astrometry and the $L'$-band photometry results from our Keck/NIRC2 imaging observations.

**Table 3.** New Keck/NIRC2 relative astrometry and $L'$ photometric measurements of the HD 176 535 AB brown dwarf–main-sequence star system.

| | HD 176 535 AB | |
|---|---|---|
| Instrument | Keck/NIRC2 | Keck/NIRC2 |
| Filter | $L'$ | $L'$ |
| Date (UT) | 2021-08-26 | 2022-07-07 |
| Epoch (Jyr) | 2021.648 75 | 2022.515 10 |
| **Relative astrometry** | | |
| Separation (mas) | $347 \pm 10$ | $362 \pm 10$ |
| PA (°) | $91.55 \pm 1.00$ | $94.81 \pm 1.00$ |
| **Photometry** | | |
| $\Delta L'$ (mag) | $9.19 \pm 0.07$ | $9.21 \pm 0.10$ |
| $L'_*$ Flux (mag) | $7.11 \pm 0.04$ | $7.11 \pm 0.04$ |
| $L'_p$ Flux (mag) | $16.30 \pm 0.07$ | $16.32 \pm 0.10$ |

*Notes.* We consistently adopt 3 per cent fractional errors in our separation and position angle measurements to aid MCMC convergence. The $L'$ magnitude for the host star is transformed from the $H$ band using Bessell & Brett (1988).

## 4 ORBITAL FIT

We determine the orbit and mass of the HD 176 535 AB system by jointly fitting the HARPS RVs, our new Keck/NIRC2 relative astrometry, and the stellar absolute astrometry from the HGCA with the Bayesian orbit fitting code ORVARA (Brandt et al. 2021b). ORVARA fits Keplerian orbits of stars and exoplanets using parallel-tempering MCMC ensemble sampler through ptemcee (Foreman-Mackey et al. 2013; Vousden, Farr & Mandel 2016). ORVARA models nine parameters for the Keplerian orbit: semimajor axis ($a$), inclination ($i$), longitude of ascending node ($\omega$), time of periastron passage ($T_0$), eccentricity and argument of periastron ($\sqrt{e}\sin\omega$ and $\sqrt{e}\cos\omega$), the masses of the two components ($M_*$ and $M_{\rm comp}$), and an RV jitter term ($\sigma_{\rm Jit}$). We ran our fit using 30 temperatures and 100 walkers over a total of $10^5$ steps to sample the nine-dimensional parameter space. We discard the first 200 000 steps as burn-in and save the coldest set of chains for parameter inferences. We adopt log-uniform priors on the companion mass, semimajor axis, and RV jitter. For the mass of the host star, we adopt a Gaussian prior of $0.72 \pm 0.02\,{\rm M}_\odot$ per our PARSEC isochrone measurement. We also test out a broader prior by inflating the uncertainty. With 3 ($0.72 \pm 0.06\,{\rm M}_\odot$) and 5 ($0.72 \pm 0.1\,{\rm M}_\odot$) times the original uncertainty, the mass posterior distributions for the companion changed to $67.9^{+2.9}_{-2.8}\,M_{\rm Jup}$ ($1.1\sigma$) and $69.3^{+3.4}_{-3.3}\,M_{\rm Jup}$ ($1.3\sigma$), respectively (see Figs A1 and A2 in the Appendix). We see a slight dependence of the companion mass on the assumed mass of the host star that reflects the covariance between companion and primary masses. There is a slight tension that we found between the observed and expected positions. This may be indication that HD 176 535 B is a binary, which could potentially account for the extra mass after error inflation and uncertainty in the astrometry. In addition, we use an informative prior on the parallax from *Gaia* EDR3 and uninformative priors on the remaining parameters.

Fig. 4 displays our orbital fit for the HD 176 535 AB system, including relative astrometry, RVs, and absolute *Hipparcos–Gaia* astrometry. The black lines highlight the maximum likelihood orbit, while the coloured lines show 50 random orbits from the MCMC posterior. The best-fitting relative astrometry orbit passes through both our relative astrometry data points.

The companion HD 176 535 B has an orbital period of $41.3^{+3.1}_{-3.6}$ yr, and the existing HARPS RV data cover around a fourth of the orbital period near its periastron passage. Therefore, as shown in the relative astrometric orbits, the constraints for the periastron part of the orbital arc are much better compared to the apastron part. Future relative astrometry beyond our 2022 data point will better constrain the semimajor axis and, as a result, the dynamical mass of the companion. The proper motions as measured by *Hipparcos* near 1991.25 and *Gaia* EDR3 near 2016.0 are the red points with error bars shown in Fig. 4. The curves reflect the astrometric reflex motion over a 25-yr time baseline between the two missions. With the newfound precision of *Gaia* EDR3, the two stellar astrometric astrometry data provide excellent constraints for this high-acceleration system.

The joint posterior distributions for important orbital parameters are showcased in Fig. 5. The posterior distributions are nearly but not perfectly Gaussian. The companion posterior mass distribution is approximately Gaussian with a value of $65.9^{+2.0}_{-1.7} M_{\text{Jup}}$ at the 68.3 per cent confidence interval. The mass is $(62.6, 70.1) M_{\text{Jup}}$ at the 95.4 per cent credible interval, which puts it definitively in the substellar regime. The HD 176 535 system has been recognized as an RV planet hosting system and an infrared excess system based on *WISE* photometry (Maldonado et al. 2017). The BD itself cannot account for the observed infrared excess: the contrast in $L'$, close to W1, is more than 9 mag. Redder *WISE* bands lie on the Rayleigh–Jeans tail of both the star's and BD's spectrum, where the contrast scales as $(R_{\text{bd}}^2 T_{\text{bd}})/(R_*^2 T_*) \sim 10^{-3}$. The source of the infrared excess remains a mystery.

The only orbital solution of this system is a recent fit by Feng et al. (2022), who found a dynamical mass of $75.5^{+8.6}_{-13.8} M_{\text{Jup}}$ for HD 176 535 B. Our dynamical mass measurement is in agreement with the lower bound of this result. The orbital fit from Feng et al. (2022) derives an RV semi-amplitude of $509^{+34}_{-123}$ m s$^{-1}$, a semimajor axis of $a = 9.1^{+1.1}_{-2.8}$ AU, an eccentricity of $0.42^{+0.04}_{-0.13}$, a period of $31^{+5}_{-13}$ yr, and an inclination of $144°.3^{+10.8}_{-8.7}$. The results are broadly consistent with our relative astrometry induced fit. In comparison, our results put the companion on a wider orbit and firmly place the companion in the BD regime, as opposed to be near the stellar/substellar boundary.

We present the first joint orbit analysis of the system, summarized in Table 4. Our best-fitting orbit suggests an inclination of $49°.8^{+3.4}_{-3.7}$, a moderate eccentricity of $0.496^{+0.022}_{-0.020}$, and a semimajor axis of $11.05^{+0.64}_{-0.56}$ au. This corroborates Bowler, Blunt & Nielsen (2020)'s finding that imaged BDs form a broad eccentricity posterior distribution with evidence for a dependence on orbital period.

## 5 DISCUSSION

### 5.1 Luminosity of HD 176535B

Here, we describe our approach to derive the luminosity for the brown dwarf HD 176 535 B from our $L'$ flux measurements. We calculate the bolometric luminosity of HD 176 535 B relying on the measured parameters for field-age (>500 Myr) ultracool substellar objects in Filippazzo et al. (2015). As shown in Fig. 7, we fit a linear relation between the absolute W1 magnitude and the bolometric magnitude estimated from integrating under the absolute flux-calibrated Spectral Energy Distribution (SED) of 15 field-age T dwarfs. We then employ the fourth-order polynomial relation between the absolute W1 magnitude and $L'$–W1 derived by Franson et al. (2023) to convert the $L'$ magnitude of HD 176 535 B to a W1 magnitude. In Section 3.3, we derived an $L'$-band contrast of $\Delta L' = 9.20 \pm 0.06$ and an apparent magnitude of $L' = 16.31 \pm 0.07$ for HD 176 535 B by combining the two measurements. We assume for the host star an $L'_* = 7.11 \pm 0.04$ as we found through the transformations of Bessell & Brett (1988). The transformation in Franson et al. (2023) gives an absolute W1 magnitude of $15.07 \pm 0.15$ for HD 176 535 B from our estimate of its $L'$ magnitude. The absolute W1 magnitude of HD 176 535 B then translates to $\log_{10}(L_{\text{bol}}/L_\odot) = -5.26 \pm 0.07$ using our derived linear relation. Here, the errors are estimated by adding the errors from the fit in Franson et al. (2023) (rms = 0.128), from our measured photometry (0.06 from the measured contrast, 0.04 from the star's $L'$ photometry), and from our $L'$–$L_{\text{bol}}$ fit (rms = 0.035) in quadrature. The bolometric luminosity we derive for HD 176 535 B matches most closely with that of Gl 229 B and $\epsilon$ Indi Bb, which places it around a spectral class of T6.

### 5.2 Model comparison

BDs of measured dynamical masses of known ages and luminosities are powerful tools for testing different evolutionary models. Our measurements of HD 176 535 B can be directly compared with the predictions from substellar evolutionary models that suggest BDs follow mass–age–luminosity relationships. The strength of the test correlates with the accuracy in mass, age, and luminosity. For instance, the binary T dwarf system $\varepsilon$ Indi B is the most precisely determined dynamical mass system that provided some of the strongest tests of substellar evolutionary models, indicating evidence of slowed cooling in the L/T transition (Chen et al. 2022).

We employ a suite of solar metallicity atmospheric and evolutionary models for cool BDs and self-luminous giant exoplanets in radiative–convective equilibrium to model the evolution of HD 176 535 B. The substellar cooling models are constructed with both an interior structural model and an exterior atmospheric model as a surface boundary condition. The model input physics, equations of states, and atmospheric boundary conditions vary across different evolutionary models for cool T-Y BDs. We consider the legacy cloudless Tucson models (Burrows et al. 1997) and the most recent ATMO-2020 models (Phillips et al. 2020). The Tucson models utilize the cloudless atmospheres of Marley et al. (1996) and the H/He equation of state from Saumon, Chabrier & van Horn (1995). The ATMO-2020 grid is also a set of cloudless evolutionary models; it is an evolution of the DUSTY, COND, and BHAC15 models (Allard et al. 2001). We adopt a hybrid model from Saumon & Marley (2008) that incorporates the effects of clouds: cloudy at $T_{\text{eff}} > 1400$ K, cloudless at $T_{\text{eff}} < 1200$ K, and a hybrid of the cloudy and cloudless atmospheres at $T_{\text{eff}}$ between 1400 and 1200 K.

We test these three BD evolutionary models by coupling any two measured parameters from mass ($M$), age ($A$), and luminosity ($L$), and compute the third from the relevant model. When the chosen parameters are mass and age, we bilinearly interpolate the evolutionary model grids by mapping the $\log(M)$–$\log(A)$ 2D coordinates to a corresponding luminosity. When the two parameters are luminosity and mass, we similarly interpolate to get age. This approach can no longer be applied when interpolating mass from an age and luminosity: mass is multiply-valued at certain ages and luminosities. We therefore construct five 2D arrays of mass as a function of age and luminosity, where five is the maximum number of values that mass can take at a given age and luminosity for any of our adopted models. We sample from each of these grids at a given age and luminosity, choosing one of the possible masses with a probability proportional to the inverse of $|dL/dM|$ at the adopted luminosity $L$ and each possible mass $M$. In all cases, we draw the two measured parameters from their posterior distributions and draw samples of the third parameter (mass, luminosity, or age) as described above.

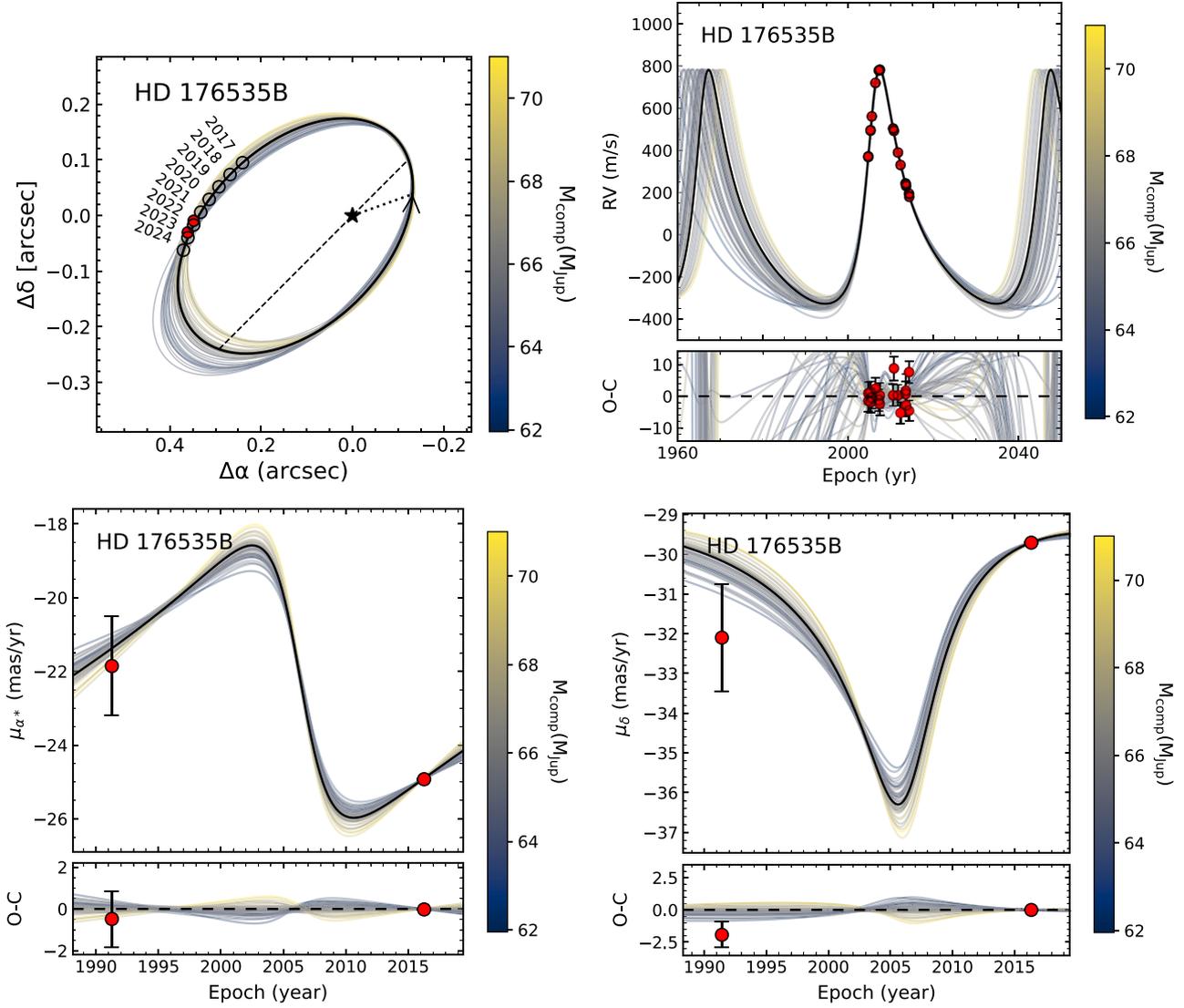

**Figure 4.** (Upper left panel) Relative astrometry from the July 2021 and August 2022 measurements is represented by the two red points. A random sampling of orbits from the MCMC steps is shown and is colour coded by the mass of the companion (from yellow to dark blue). The black orbit is the best-fitting orbit; its $\chi^2$ value indicates a formally a good fit. (Upper right panel) The RV orbit of the companion. The random draws from the MCMC posterior fit tightly to the RVs along with the best-fitting orbit. The periastron part of the orbit is well constrained compared to the apastron part of the orbit. (Bottom panels) Proper motions from the model compared to the calibrated *Hipparcos* and *Gaia* EDR3 proper motions near 1991.25 and 2016, respectively. Both *Hipparcos* and *Gaia* EDR3 proper motions constrain the proper motion of the star well, *Gaia* EDR3 especially. The integrated proper motion between the *Hipparcos* and *Gaia* points is constrained by the long-term HGCA proper motion.

Fig. 6 shows $A$–$L$ and $M$–$L$ relations from the three evolutionary models of our choices, as well as our measured benchmark BD HD 176 535 B sitting atop the iso-age and iso-mass evolutionary grids. The TUCSON and ATMO2020 models are consistent with our photometry, age and mass measurements of HD 176 535 A within $3\sigma$ uncertainties, while the hybrid model prefers a much older age and/or lower masses for the companion. The discrepancy between the models and our measurement indicates possible tension with evolutionary models.

Fig. 8 illustrates the inferred mass and age distributions compared to the posterior distributions for HD 176 535 B. We compute the discrepancy between our model-derived distributions and model-independent distributions by randomly drawing ages or masses from both distributions to determine what fraction of data is one value larger than the other. We then express this fraction in units of sigma via a per cent point function or the inverse of the cumulative distribution function of the Gaussian, i.e. probability to the left of the distribution. We find that our dynamical mass distribution is consistent with the TUCSON and ATMO2020 models within $3\sigma$, but slightly disagree with model predictions from the SM08 hybrid model. The model-predicted ages deviate from our derived age by $1.7\sigma$ (TUSCON model), $2.4\sigma$ (ATMO 2020), and $3.0\sigma$ (SM08 Hybrid). The model-derived mass distributions deviate from our MCMC posterior masses by $2.0\sigma$ (TUCSON), $2.9\sigma$ (ATMO 2020), and $3.5\sigma$ (SM08 Hybrid), respectively. Both the cloudless TUCSON and ATMO2020 models predict similar, albeit broader mass distributions compared to the MCMC mass posterior distribution. The Saumon & Marley (2008) hybrid model predicted a distribution much lower than our results from MCMC. Our results suggest that substellar evolutionary models may be systematically underestimating luminosity for high-mass T dwarfs; however, improved age estimates will help to better constrain the mass predictions from all models.

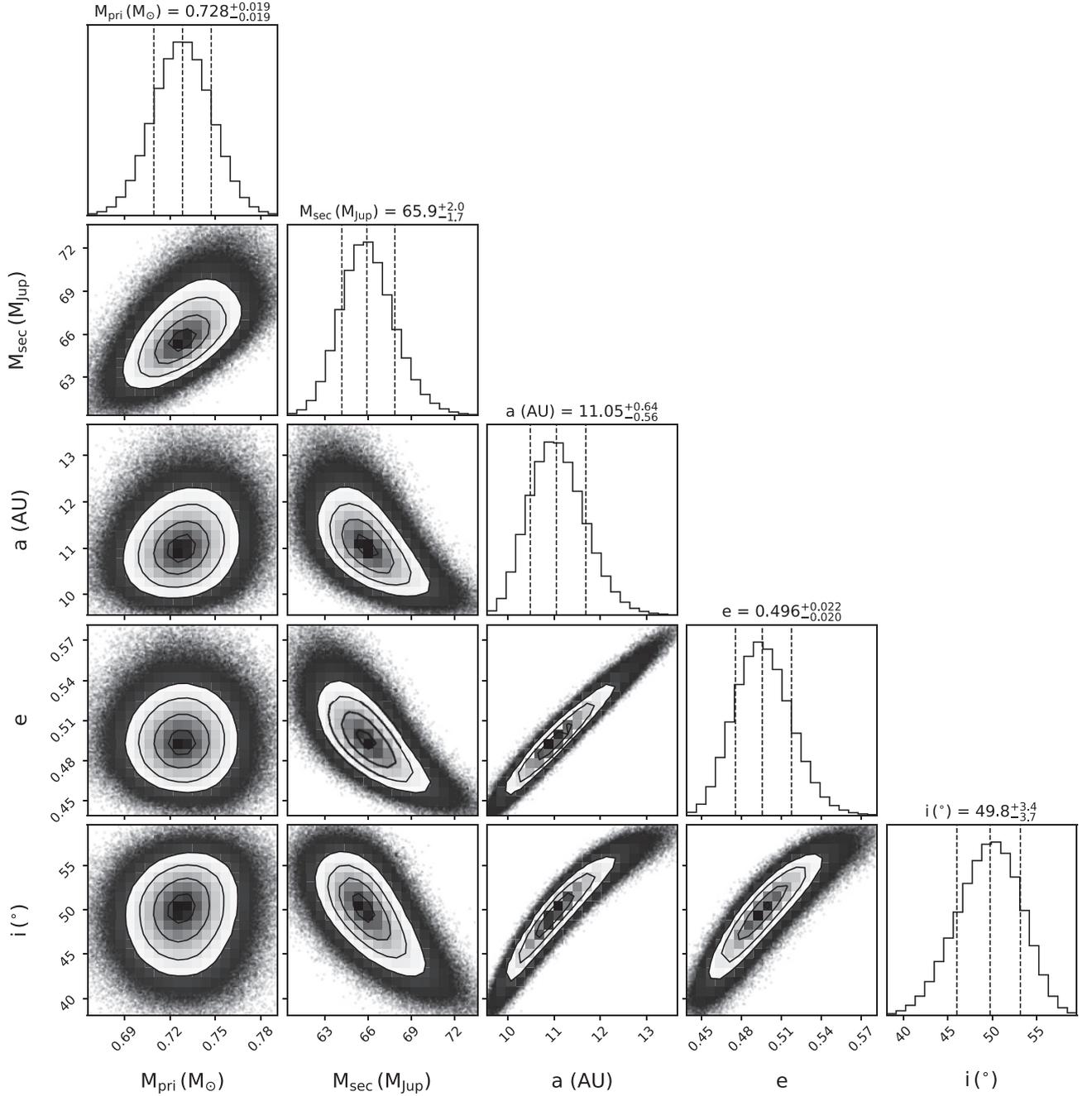

**Figure 5.** Joint posterior distributions for selected parameters for HD 176 535 B. The errors are in the 16 and 84 per cent quantiles about the median. The 2D contours give the $1\sigma$, $2\sigma$, and $3\sigma$ levels that show the correlations between any two parameters.

Benchmark BDs with measured spectra may be used to test model atmospheres. Our bolometric luminosity and effective temperature measurements can be used against the evolutionary models to see how well they agree with the models. The BD radii as a function of mass, age, and metallicity predicted by the evolutionary models can be combined with empirically determined luminosity to produce independent estimates of $T_{\rm eff} = (L_{\rm bol}/4\pi R^2 \sigma_{\rm SB})^{1/4}$ and $\log(g) = \log(GM/R^2)$, where $\sigma_{\rm SB}$ is the Stefan–Boltzmann constant.

Table 5 documents the derived parameters from each model and our measured values in comparison. Our model-derived temperatures of $\approx$1000 K indicate that HD 176 535 B is a massive, late T-type BD compared to the observed SED sample from Filippazzo et al. (2015).

**Table 4.** MCMC orbital fit results for HD 176 535 AB.

| Parameter | Prior | Best fit | 68.3 per cent CI | 95.4 per cent CI |
|---|---|---|---|---|
| | | Fitted parameters | | |
| $\sigma_{\rm Jit}$ (m s$^{-1}$) | $1/\sigma_{\rm Jit}$ | 3.83 | $3.83^{+1.1}_{-0.82}$ | (2.40, 6.47) |
| $M_*$ (M$_\odot$) | $N(0.72, 0.02)$ | 0.728 | $0.728^{+0.019}_{-0.019}$ | (0.691, 0.766) |
| $M_{\rm p}$ (M$_\odot$) | $1/M_{\rm p}$ | 65.9 | $65.9^{+2.0}_{-1.7}$ | (62.6, 70.1) |
| $a$ (au) | $1/a$ | 11.05 | $11.05^{+0.64}_{-0.56}$ | (10.03, 12.38) |
| $\sqrt{e}\sin\omega$ | $U(-1, 1)$ | $-0.373$ | $-0.373^{+0.014}_{-0.012}$ | $(-0.398, -0.344)$ |
| $\sqrt{e}\cos\omega$ | $U(-1, 1)$ | 0.597 | $0.597^{+0.023}_{-0.021}$ | (0.557, 0.644) |
| $i$ (°) | $\sin i$ | 49.8 | $49.8^{+3.4}_{-3.7}$ | (42.4, 56.1) |
| $\Omega$ (°) | $U(-180, 180)$ | 129.51 | $129.51^{+0.94}_{-1.2}$ | (126.8, 131.2) |
| $\lambda_{\rm ref}$ (°) | $U(-180, 180)$ | $-176.9$ | $-176.9^{+356}_{-2.3}$ | $(-179.885, 179.883)$ |
| | | Derived parameters | | |
| $\varpi$ (mas) | – | 27.0326 | $27.0326^{+0.0011}_{-0.0011}$ | (27.03, 27.035) |
| $P$ (yr) | – | 41.3 | $41.3^{+3.6}_{-3.1}$ | (35.7, 48.9) |
| $\omega$ (°) | – | 328.0 | $328.0^{+1.7}_{-1.6}$ | (324.8, 331.6) |
| $e$ | – | 0.496 | $0.496^{+0.022}_{-0.020}$ | (0.459, 0.541) |
| $a$ (mas) | – | 299 | $299^{+17}_{-15}$ | (271, 335) |
| $T_0$ | – | 2468 925 | $2468\,900^{+1300}_{-1100}$ | (2466 900, 2471 700) |

*Note.* The reference epoch is 2455197.5 JD.

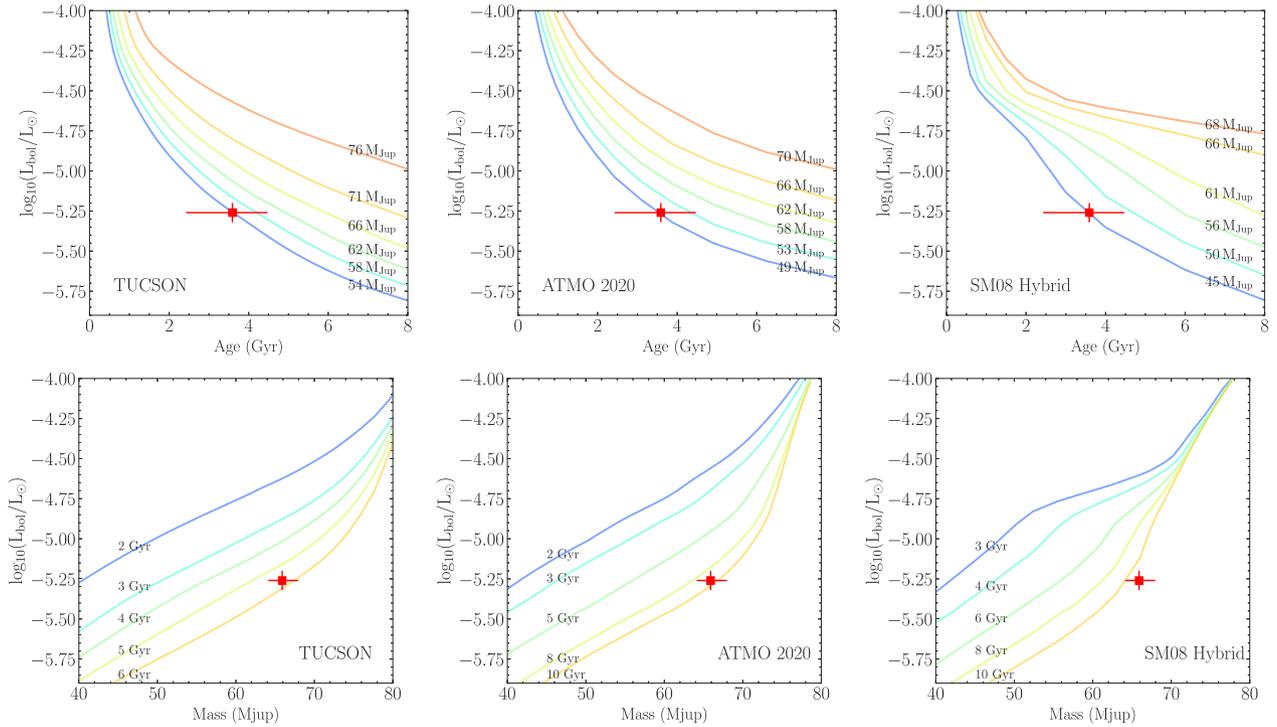

**Figure 6.** Different evolutionary models compared with the astrometric and photometric measurements of the brown dwarf HD 176 535 B. Three independent substellar cooling models are considered: TUCSON, ATMO2020, and the SM08 hybrid model. The substellar cooling curves are the iso-mass and iso-age lines derived from each model given the mass and age of HD 176 535 B. The top panels show the bolometric luminosity against substellar cooling age, and bottom panels show the bolometric luminosity against the mass of the BD.

### 5.3 The T dwarf population

Our measurement for HD 176 535 B is comparable to that of several other massive (60–75 $M_{\rm Jup}$) benchmark T dwarfs of intermediate age (1–6 Gyr) that are slightly inconsistent with evolutionary model predictions of luminosity as a function of dynamical mass. HD 176 535 B is overmassive given our measurements for its age and luminosity. Here, we discuss the case of HD 176 535 B in the context of other well-measured benchmark L/T- or T dwarf systems (see comprehensive lists in Dupuy & Liu 2017; Brandt et al. 2021d; Franson et al. 2022, and discussions within). We plot the luminosities of these systems against their masses, together with the evolutionary grids from the Saumon & Marley (2008) hybrid model in Fig. 9. All of these T dwarfs are directly imaged, and have orbits constrained from both RV and *Gaia* DR2 parallaxes or newer *Gaia* EDR3 parallaxes in the cases of Gl 758 B, HD 13 724 B, HD 19 467 B, HD 33 632 Ab, HD 72 946 B (Brandt et al. 2021d), and the binary BDs $\varepsilon$ Indi Ba and Bb (Chen et al. 2022).

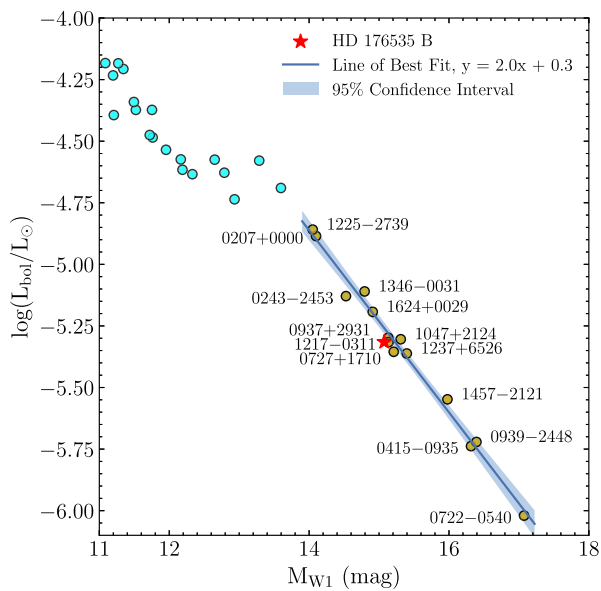

**Figure 7.** W1 as a function of $L_{bol}$ using the field BD SED sample from Filippazzo et al. (2015). The cyan dots are L dwarfs, and the orange dots are field T dwarfs that we fit a linear relation to. We derive a relation between the absolute W1 magnitude of the T dwarfs and the log of their luminosity as $\log(L_{bol}/L_\odot) = 2.0 M_{W1} + 0.3$. The red star places HD 176 535 B among the T dwarf sequence in the SED sample, and we infer a spectral type of T6 for HD 176 535 B.

The benchmark BDs of T or L/T transition types shown here include Gl 802 B ($66 \pm 5 M_{Jup}$, aged $\geq 1$ Gyr; Ireland et al. 2008), HD 4747 B ($66.2 \pm 2.5 M_{Jup}$, aged $3.6 \pm 0.6$ Gyr; Brandt et al. 2019), $\varepsilon$ Indi Ba (a $66.96 \pm 0.35 M_{Jup}$ L/T transition BD aged $3.5^{+0.8}_{-1.0}$ Gyr; Chen et al. 2022), HD 72 946 B (a massive T dwarf $72.5 \pm 1.3 M_{Jup}$ aged $5.4 \pm 1.9$ Gyr; Brandt et al. 2021d), and HR 7672 B ($72.7 \pm 0.8 M_{Jup}$, aged $5.4 \pm 1.9$ Gyr). Although most of these benchmark BDs are in agreement with the hybrid evolutionary models, there are a couple of outliers that point to a tension in the consistency of evolutionary models. The ones that are consistent with the models within $3\sigma$ given their masses and ages include HD 72 946 B, HR 7672 B, HD 4747 B, HD 33 632 Ab, $\varepsilon$ Indi Ba/Bb, and Gl 758 B. Being the only system in this list with two T dwarfs of the same age, the $\varepsilon$ Indi B binary system provided a unique and strong test for evolutionary models (Chen et al. 2022). The very different luminosities of the two BDs and the moderate mass ratio between the two suggest a steep $M$–$L$ relationship $L \propto M^{5.47 \pm 0.08}$ that can be explained by a slowed cooling rate in the L/T transition.

However, despite the fact that many T dwarfs are consistent with the models, there is a non-trivial fraction of overluminous (undermassive, overaged) and underluminous (overmassive, underaged) BDs that challenge the substellar evolutionary models' range of validity. The overluminous BDs are brighter than what models predict given their independent mass and age measurements, e.g. HD 13 724 B (older than expected from substellar cooling ages) (Rickman et al. 2020), HD 130 948 BC (Geißler et al. 2009), Gl 417 BC (Dupuy, Liu & Ireland 2014), and CWW 89 Ab (Beatty et al. 2018). The underluminous cases mean that the measured luminosities are fainter than model predictions given their masses and ages. These include the first discovered T dwarf Gl 229 B (Calamari et al. 2022), HD 19 467 B (Brandt et al. 2021d), HD 47 127 B (Bowler et al. 2021b), and HD 4113 C (Cheetham et al. 2018), and our BD HD 176 535 B.

Our measurement of HD 176 535 B joins it among a growing list of slightly overmassive and underluminous benchmark BDs. According to our derivation of HD 176 535 B's bolometric luminosity, the most immediate analogues for HD 176 535 B are HD 19 467 B, Gl 229 B, and $\varepsilon$ Indi Bb (see Fig. 9). The closest match is HD 19 467 B, a massive ($65 \pm 6 M_{Jup}$) T dwarf (T5.5 $\pm$ 1) near the substellar mass boundary and aged $5.4 \pm 1.9$ Gyr on an eccentric orbit around a G3V star (Crepp et al. 2015; Brandt et al. 2021d). All of these analogues have spectral types of T6/T7, which indicates that HD 176 535 B is likely also a T6 dwarf – its spectral type can/will further be confirmed and characterized with the CHARIS spectrograph (Lewis et al., in preparation). While the analogues of HD 176 535 B and itself have similar luminosity, they differ in terms of agreeing/disagreeing with evolutionary models. Namely, $\varepsilon$ Indi Bb agrees with substellar evolutionary models (Chen et al. 2022), Gl 229 B strongly disagrees with evolutionary models, and both HD 176 535 B and HD 19 467 B somewhat disagree with the models, but to a lesser extent than Gl 229 B. Gl 229 B, with a mass of $71.4 \pm 0.6 M_{Jup}$ and an age of $5.4 \pm 1.9$ Gyr, is anomalously overmassive and way too faint for its mass at the model-predicted substellar cooling ages (Filippazzo et al. 2015; Brandt et al. 2020, 2021d; Calamari et al. 2022). The reason for the different degrees of discrepancy between measurements and models we see in these BDs may arise from unresolved binarity of the systems. It is plausible that the Gl 229 B system has an unresolved high-mass companion, but HD 176 535 B and HD 19 467 B also have unresolved, albeit low-mass ratio companions. Unfortunately, in all cases, unresolved companion would be too faint to be detected in multidecade high-resolution spectra (Brandt et al. 2020). Therefore, the origin of HD 176 535 B's relatively high mass is ambiguous, but we cannot rule out binarity as an explanation. Other likely explanations include the following: (1) We are substantially underestimating ages of T dwarfs; for example, the ages of HD 176 535 B and HD 19 467 B are actually closer to 8–10 Gyr as opposed to 3–5 Gyr. (2) Missing physics in the BD cooling models.

In this study, we compared HD 176 535 B with three models that are predominantly cloud-free given the object's temperature and the hybrid model's degeneracy in cloud treatment. A more accurate estimate of the bolometric correction would require using a range of atmosphere models that account for cloud effects. Unfortunately, no other evolutionary models presently available integrate synthetic spectra with evolutionary tracks, rendering it impracticable to correlate these models with HD 176 535 B's mass, luminosity, and age. Thus, we resorted to an empirical correction approach that merely necessitates a single-band measurement in $L'$. Future multiwavelength measurements will provide a more precise bolometric correction.

Till now, it is still unclear whether the overmassive T dwarfs are just a handful of unresolved binaries or a more severe systematic problem for the T dwarf cooling models. As more L, T, and L/T transition BDs are being discovered and characterized, a complete picture on the evolution and cooling of BDs across the L, T, and Y spectra will be better portrayed.

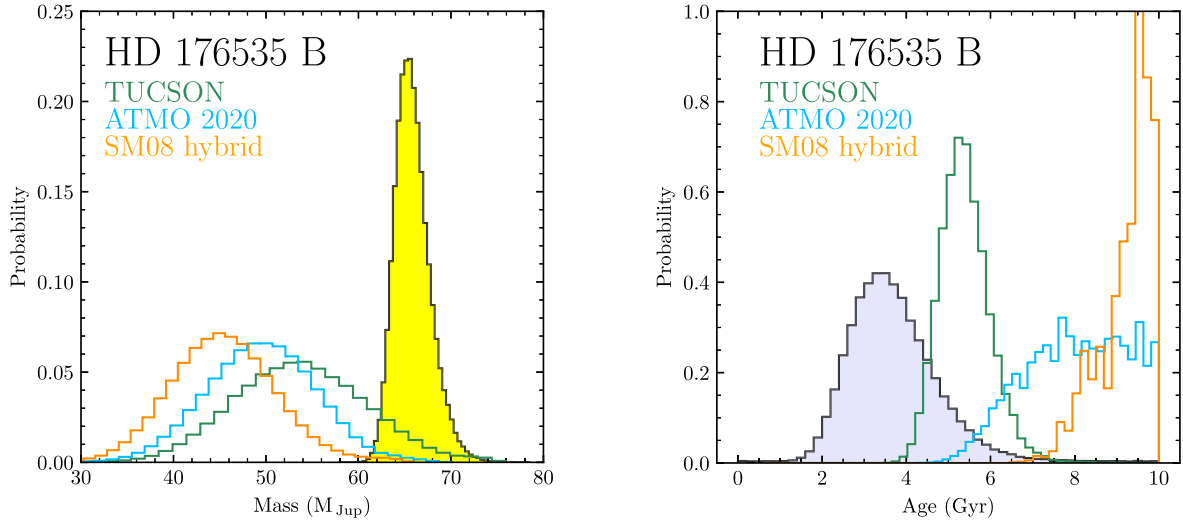

**Figure 8.** (Left) Dynamical mass posterior distribution (filled histogram) compared to the model-derived mass posteriors given $L_{bol}$ and host star ages (step histograms). (Right) Host star activity-based age probability distribution (filled histogram) compared to the model-derived age posteriors given $L_{bol}$ and mass. All three models produce model-derived results that are consistent with the measured MCMC mass posteriors and isochronal age distributions.

**Table 5.** Evolutionary model-derived parameters versus measured parameters.

| Property | TUCSON | ATMO2020 | SM08 hybrid | Measurement |
|---|---|---|---|---|
| Mass ($M_{Jup}$) | $52.9^{+7.1}_{-7.3}$ ($2.0\sigma$) | $48.8^{+6.0}_{-5.8}$ ($2.9\sigma$) | $44.6^{+5.5}_{-5.7}$ ($3.5\sigma$) | $65.9^{+2.0}_{-1.7}$ |
| $\log(L_{bol}/L_\odot)$ | $-4.95 \pm 0.20$ ($1.8\sigma$) | $-4.77 \pm 0.19$ ($2.7\sigma$) | $-4.63 \pm 0.11$ ($3.3\sigma$) | $-5.26 \pm 0.07$ |
| Age (Gyr) | $5.25^{+0.54}_{-0.61}$ ($1.7\sigma$) | $8.03^{+1.42}_{-1.23}$ ($2.4\sigma$) | $9.35^{+0.91}_{-0.34}$ ($3.0\sigma$) | $3.59^{+0.87}_{-1.15}$ |
| $T_{eff}$ (K) | $982 \pm 34$ | $988 \pm 35$ | $950 \pm 36$ | – |
| $\log(g)$ (cm s$^{-2}$) | $5.32 \pm 0.06$ | $5.30 \pm 0.06$ | $5.19 \pm 0.06$ | – |
| Radius ($R_{Jup}$) | $0.787^{+0.010}_{-0.009}$ | $0.776^{+0.012}_{-0.010}$ | $0.840^{+0.023}_{-0.024}$ | – |

*Notes.* Masses are computed using the measured $\log(L_{bol})$ and age; $\log(L_{bol})$ from measured mass and age; and age from measured mass and $\log(L_{bol})$. Model-derived radii use the measured age and mass; and $T_{eff}$ and $\log g$ use model radii and measured mass.

### 5.4 Future follow-ups and *Gaia* DR4 accelerations

We have carried out direct imaging measurements of HD 176 535 B in the $L'$ band with the NIRC2/Keck AO pyramid WFS. HD 176 535 B turned out to be a massive, T-type dwarf according to our ORVARA orbital fit. The mass and eccentricity of the system are very well constrained from our orbital fit to less than 3 per cent. Any future high-contrast imaging measurements will not only help refine the mass and orbital parameters, but also determine the spectral type of HD 176 535 B. Multiband near-infrared spectra with instruments such as the CHARIS/SCExAO/Subaru and the Keck Planet Imager and Characterizer (KPIC/NIRSPEC), as well as photometry from *JWST* in the thermal infrared, will enable spectroscopic characterizations of HD 176 535 B. Further RV monitoring beyond 2024 could also enhance the orbital fit, especially the orbit when the companion is near apastron. In addition to RVs, future *Gaia* epoch astrometry such as *Gaia* DR4 and DR5 will slowly resolve HD 176 535 B's astrometric orbit as well (see bottom two figures of Fig. 4).

Given HD 176 535 B's current ORVARA orbit presented in this paper, we predict the star HD 176 535 A's DR4 accelerations using an adaptation of the epoch astrometry code HTOF (Brandt et al. 2021c). This adaptation parses the intermediate *Gaia* scanning law data from GOST from 2014.6403 to 2020.1403 (this covers the 5.5-yr baseline for DR4), and densely samples the orbits to fit astrometric solutions to those data. The predicted *Gaia* DR4 accelerations are summarized in Table 6. These accelerations can be directly compared to the DR4 accelerations when they are released. These accelerations provide a unique opportunity, along with 62 other predictions in An et al. (in preparation), to test the accuracy of the *Gaia* data. These accelerating systems act as 'dynamical beacons' that anchor *Gaia* astrometry for non-single stars.

### 6 CONCLUSION

In this work, we measured the dynamical mass of the imaged substellar companion HD 176 535 B discovered in the first phase of a pilot discovery programme targeting accelerating stars using the NIRC2 camera at the WMKO. The companion's on-sky location was predicted in advance of imaging with the orbit-fitting package ORVARA to be a $62M_{Jup}$ companion on a wide orbit. It was subsequently observed for the first time with Keck/NIRC2 in the $L'$ band in August 2021, and followed up in July 2022. After our observations, we carried out orbital fits with the dual-epoch relative astrometry, RVs from HARPS, and *Gaia* EDR3 absolute astrometry from the HGCA. Our model-independent dynamical mass measurement of $65.9^{+2.0}_{-1.7} M_{Jup}$ revealed an old ($3.59^{+0.87}_{-1.15}$ Gyr) and massive BD companion. The thermal spectra of the companion have not been measured yet, but model-dependent measurements of effective temperature and surface gravity place it in the T dwarf regime. Future astrometric and spectroscopic measurements (e.g. in the $H$ and $K$ bands with CHARIS/SCExAO/Subaru) confirm its T-class spectral type. HD 176 535 B joins a growing list of benchmark BDs that have precisely determined dynamical masses, ages, and luminosities. Future data from radial velocities, direct imaging relative astrometry, and/or *Gaia* epoch astrometry will place better constraints for its orbit. Finally, multiband infrared spectroscopy follow-up will be useful to identify its spectra and to calibrate substellar evolutionary models. The successful imaging of HD 176 535 B following its prediction is a proof of the potential for informed direct imaging searches. HGCA-based imaging surveys across the sky are revolutionary, and will continue to discover more faint companions with significant accelerations until the orbits for all the nearby BDs and giant exoplanets are fully characterized.

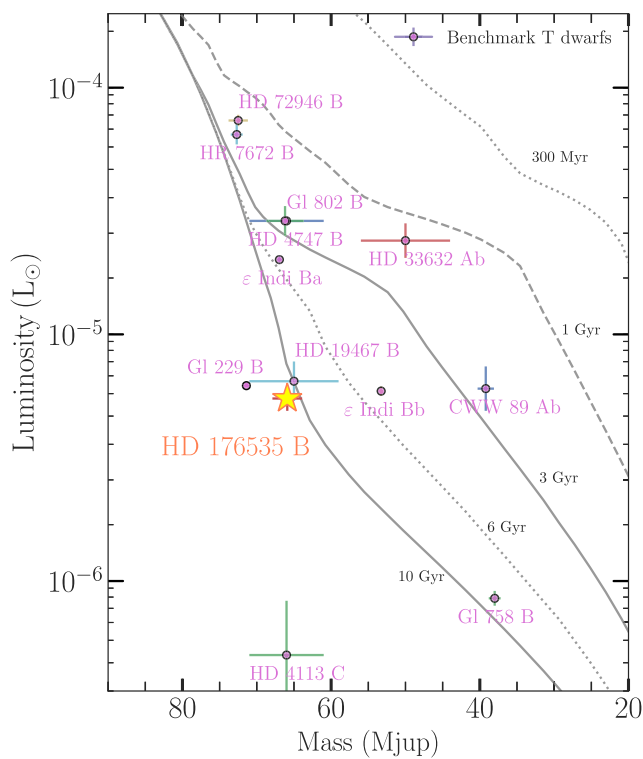

**Figure 9.** Luminosity as a function of mass for literature benchmark T dwarfs (purple), our target HD 176 535 B (coral), as well as the hybrid Saumon & Marley (2008) evolutionary models (grey lines). Note that HD 4747 B, CWW 89Ab, $\varepsilon$ Indi Ba, and HD 33 632 Ab are L/T transitioning BDs. The error bars on the binary BDs $\varepsilon$ Indi Ba and Bb are the tightest because they have the most precisely measured masses up to date. The literature masses and luminosities are from Dupuy & Liu (2017), Dupuy et al. (2019), Brandt et al. (2021d), Chen et al. (2022), and Franson et al. (2022).

**Table 6.** Predicted *Gaia* DR4 accelerations at epoch = 2017.5.

| Parameter | Value (mas yr$^{-2}$) |
|---|---|
| Predicted $\dot{\mu}_{\alpha*}$ | $0.223 \pm 0.0323$ |
| Predicted $\dot{\mu}_{\delta}$ | $0.059 \pm 0.0102$ |

## ACKNOWLEDGEMENTS

This research has made use of the KOA, which is operated by the WMKO and the NExScI, under contract with the National Aeronautics and Space Administration. This work presents results from the European Space Agency (ESA) space mission *Gaia*. *Gaia* data are being processed by the Gaia Data Processing and Analysis Consortium (DPAC). Funding for the DPAC is provided by national institutions, in particular the institutions participating in the *Gaia* MultiLateral Agreement (MLA). The *Gaia* mission website is https://www.cosmos.esa.int/gaia. The *Gaia* archive website is https://archives.esac.esa.int/gaia. TDB gratefully acknowledges support from the Alfred P. Sloan Foundation and from the NASA Exoplanet Research Program under grant #80NSSC18K0439. BLL acknowledges support from the National Science Foundation Graduate Research Fellowship under Grant No. 2021-25 DGE-2034835. Any opinions, findings, and conclusions or recommendations expressed in this material are those of the authors(s) and do not necessarily reflect the views of the National Science Foundation. BPB acknowledges support from the NASA Exoplanet Research Program grant 20-XRP20 2-0119 and the Alfred P. Sloan Foundation. RK acknowledges the support by the National Science Foundation under Grant No. NSF PHY-1748958. We thank the Heising Simons Foundation for the support.

## DATA AVAILABILITY

This paper includes data collected by the NIRC2 camera at the WMKO, which is publicly available from the Keck Observatory Archive (KOA) jointly managed by the WMKO and NASA Exoplanet Science Institute (NExScI). Funding for this research is provided by the National Aeronautics and Space Administration. All data in this paper are publicly available through the KOA, except for the most recent data for July 2022. The data are analysed with the ORVARA and VIP open-source packages, which are publicly available at https://github.com/t-brandt/orvara and https://github.com/vortex-exoplanet/VIP, respectively. We acknowledge the use of public *Gaia* EDR3 data through the *Gaia* archive (https://gea.esac.esa.int/archive/).

# APPENDIX A

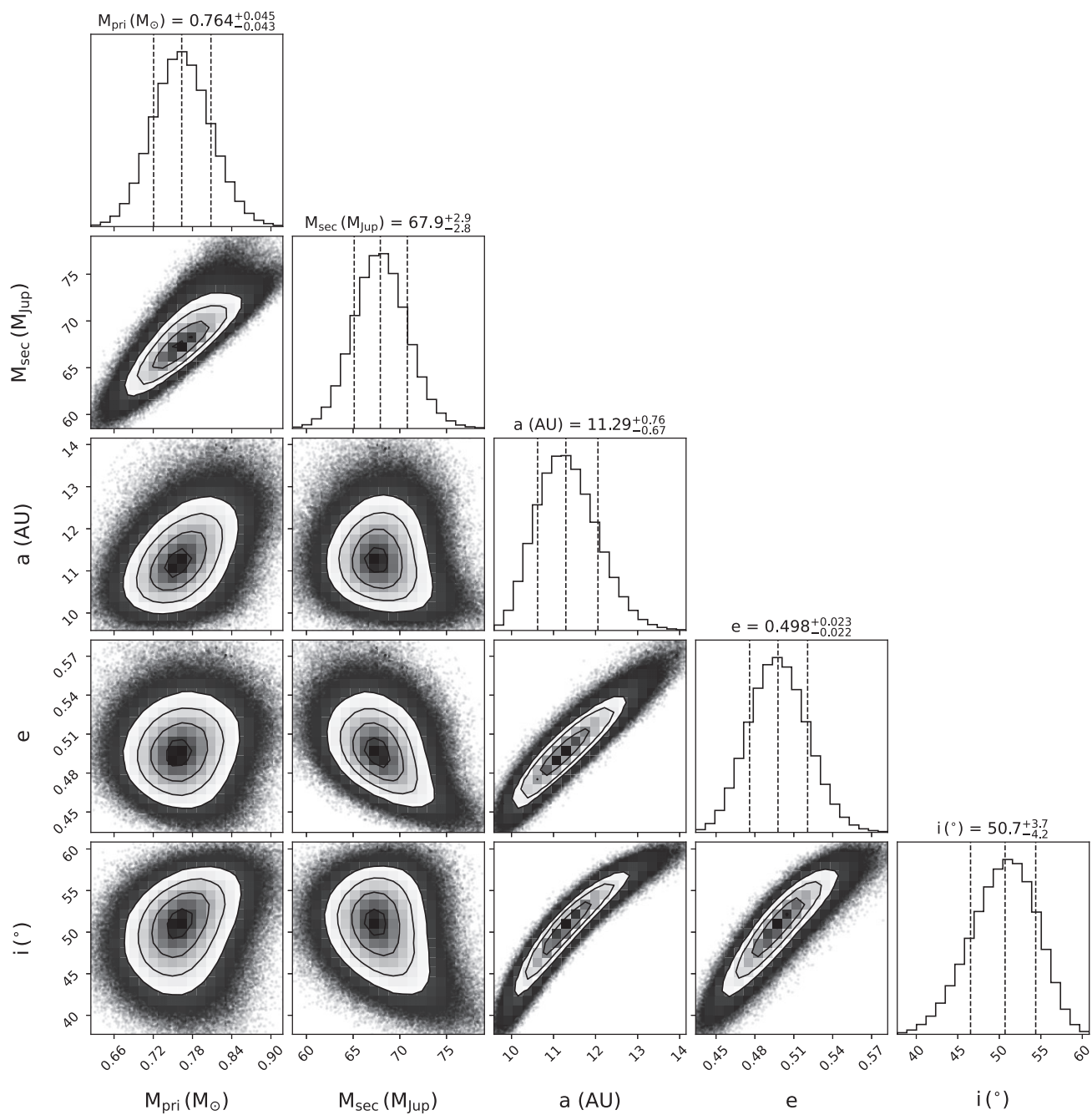

**Figure A1.** Joint posterior distributions for selected parameters for HD 176535 B with a prior of $0.72 \pm 0.06\,\mathrm{M}_\odot$ on the host star. The errors are in the 16 and 84 per cent quantiles about the median. The 2D contours give the $1\sigma$, $2\sigma$, and $3\sigma$ levels that show the correlations between any two parameters.

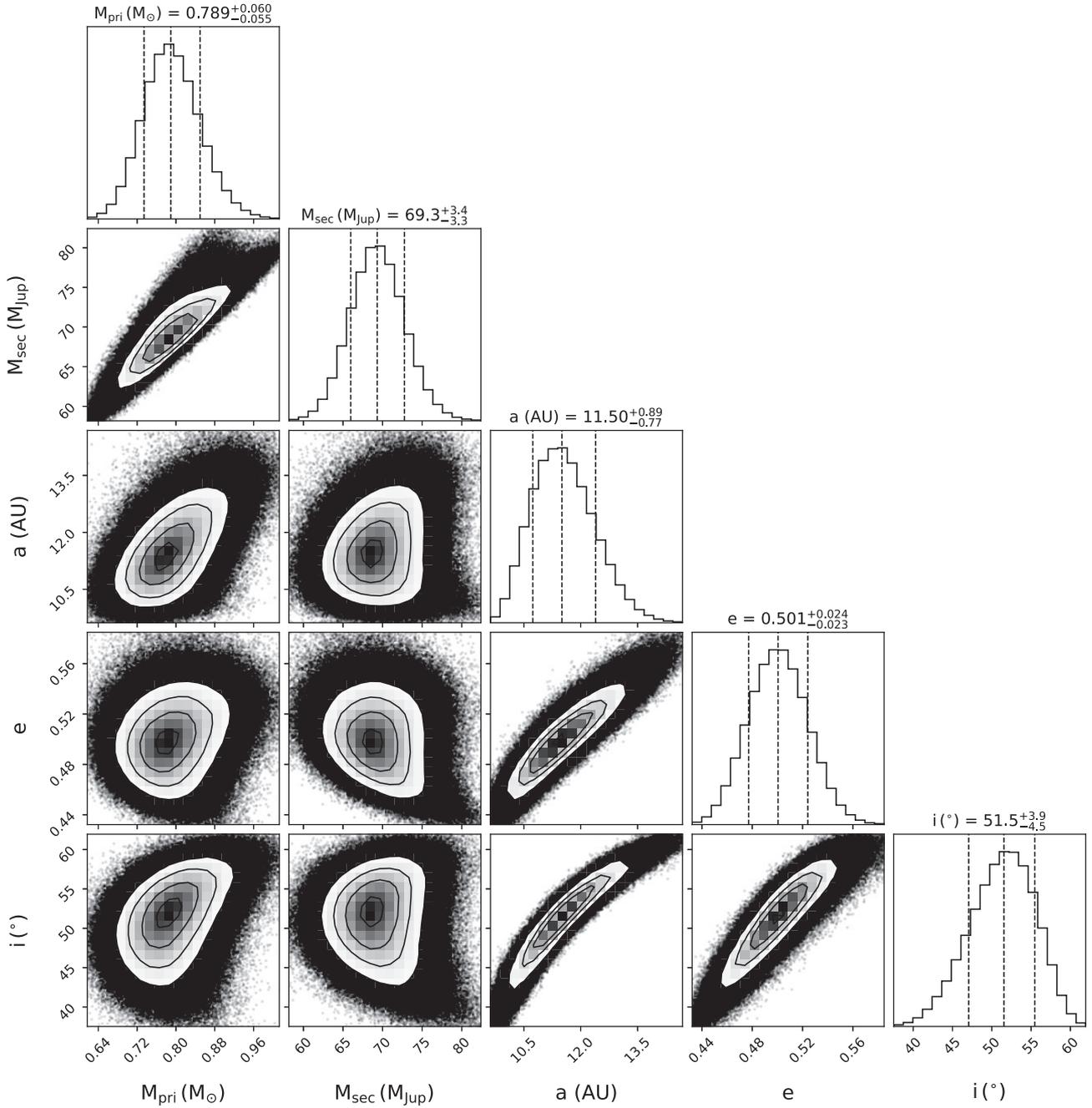

**Figure A2.** Joint posterior distributions for selected parameters for HD 176535 B with a prior of $0.72 \pm 0.1\,M_\odot$ on the host star. The errors are in the 16 and 84 per cent quantiles about the median. The 2D contours give the $1\sigma$, $2\sigma$, and $3\sigma$ levels that show the correlations between any two parameters.

This paper has been typeset from a TEX/LATEX file prepared by the author.